\documentclass[twocolumn]{aastex6}

\usepackage{epstopdf}
\usepackage{amsmath}
\usepackage{amssymb}
\usepackage{graphicx}

\DeclareMathAlphabet      {\mathbfit}{OML}{cmm}{b}{it}

\shorttitle{Coronal Dynamic Activities}
\shortauthors{Jang et al.}

\begin{document}


\title{Coronal Dynamic Activities in the Declining Phase of a Solar Cycle}



\author{ Minhwan~Jang\altaffilmark{1, 2}, 
T.~N.~Woods\altaffilmark{3}, 
Sunhak~Hong\altaffilmark{4, 2}, 
and
G.~S.~Choe\altaffilmark{1, 2} }



\altaffiltext{1}{Department of Astronomy and Space Science, Kyung Hee University, Yongin 17104, Korea; gchoe@khu.ac.kr}
\altaffiltext{2}{School of Space Research, Kyung Hee University, Yongin 17104, Korea}
\altaffiltext{3}{Laboratory for Atmospheric and Space Physics, University of Colorado, Boulder, CO 80303, USA}
\altaffiltext{4}{Korean Space Weather Center, National Radio Research Agency, Jeju 63025, Korea}


\begin{abstract}

It has been known that some solar activity indicators show a double-peak feature in their evolution through a solar cycle, which is not conspicuous in sunspot number.  
In this Letter, we investigate the high solar dynamic activity in the declining phase of the sunspot cycle by examining the evolution of polar and low-latitude coronal hole (CH) areas, splitting and merging events of CHs, and coronal mass ejections (CMEs) detected by SOHO/LASCO C3 in solar cycle 23.  Although the total CH area is at its maximum near the sunspot minimum, in which polar CHs prevail, it shows a comparable second maximum in the declining phase of the cycle, in which low latitude coronal holes are dominant. The events of CH splitting or merging, which are attributed to surface motions of magnetic fluxes, are also mostly populated in the declining phase of the cycle. The far-reaching C3 CMEs are also overpopulated in the declining phase of the cycle. From these results we suggest that solar dynamic activities due to the horizontal surface motions of magnetic fluxes extend far in the declining phase of the sunspot cycle.

\end{abstract}





\section{Introduction} \label{sec:intro}

The sun shows a periodic behavior in its activity, which includes formation of sunspots and occurrence of solar eruptive phenomena such as solar flares and coronal mass ejections (CMEs).  The periodic variation of sunspot numbers was first discovered by \citet{Schwabe1844}, and \citet{Wolf1861} subsequently devised a quantity to measure the sunspot activity, the so-called Wolf, Z{\" u}rich, or International sunspot number. Since \citet{Hale1908} measured the magnetic field in sunspots, the sunspot number has been regarded as a proxy of solar magnetic activity. 
The smoothed sunspot number (SSN) as a function of time generally has one minimum and one maximum during  an 11 year solar cycle. However, the sunspot number in a solar cycle often shows double-peak (maxima) features as in the recent solar cycles 22, 23, and 24. These double-peak trends of solar activities are more conspicuous in coronal emission as noted by \citet{Gnevy1963, Gnevy1967, Gnevy1977} than in sunspot number, and the dipped period of 1--2 years between the two maxima in various solar activities is called the Gnevyshev gap \citep{Storini1995, Kane2005}. The definition and length of the Gnevyshev gap widely vary depending on which solar activity indicator is referenced. 

There have been quite a few studies on the secondary variability of solar activities other than sunspot generation \citep[e.g.,][]{Du2015, McIntosh2015}. However, not all solar activity indicators show a double-peak feature. In a comprehensive review on the solar cycle, \citet{Hathaway2015} has compared SSN with other measurable quantities representing solar activities, for example, diverse sunspot numbers, sunspot areas, the solar radio flux, the total solar irradiance, solar flare numbers, and the geomagnetic AA index. Most of these indicators show a very good correlation with SSN. In particular, the frequency of M- and X-class flare occurrence quite well follows the time evolution of SSN with only a 2 month time lag, which is far shorter than a usual Gnevyshev gap. The geomagnetic AA index, however, evidently shows large values in the declining phase of the sunspot cycle, which is attributed to the movement of coronal holes (CH) toward lower latitudes after the solar maximum \citep{Legrand1985}. 


In this Letter, we investigate the solar ``dynamic'' activity trend in relation to the CH evolution in solar cycle 23. First, we examine the time variation of CH areas. The CH area can be a rough proxy of the open magnetic flux of the sun. Although there are more sophisticated ways of estimating the open magnetic flux of the sun \citep{Lockwood1999, Solanki2000, Solanki2002}, the CH area can almost directly be measured with a higher time cadence than the quantities used in those methods.  
In a comprehensive study based on satellite observations for 40 years,  \citet{Zhou2009} showed that the heliospheric magnetic flux varies up to a factor of 2 over a solar cycle.  This observation leads us to take   a source surface model of the global coronal magnetic field \citep{Altschuler1969, Schatten1969, Wang1992} as a theoretical framework for interpretation of our data because the model accommodates variation of the open flux.
Our study examines the polar coronal area and the area of low latitude CHs discriminately. By doing this, we can guess whether a dipole moment or higher-order multipole moments are dominant in the global solar magnetic field at a certain time. 


Based on the potential field source surface (PFSS) model, \citet{Antiochos2007} argued that there should be only one CH in each unipolar region. 
However, \citet{Titov2011} have shown that CHs of the same magnetic polarity can be connected by singular lines of zero width. Thus, CHs of one polarity can be observed as separate entities. This fact has been confirmed in an extensive PFSS-based study by \citet{Platten2014}. \citet{Titov2011} have also shown that a motion of magnetic patches can either split a CH or merge separate CHs. We pay attention to this reconfiguration of CHs because such events are unarguable evidences of a very dynamic stage in the field footpoint regions. This Letter, for the first time, addresses the statistics of splitting and merging events of CHs. 

It is somewhat puzzling how the open magnetic flux can globally increase or decrease \citep{Fisk2001, Crooker2002, Owens2006, Owens2011}. In many CME observations, the CME structure exits the field of view of the coronagraph and the remaining field lines look open with respect to the outer boundary of the coronagraph \citep{Hundhausen1999}. After the ``field opening,'' magnetic reconnection is expected to take place in the resulting current sheet, producing a flare to recover the closed field structure. However, not all the opened flux is likely to reconnect again because a substantial magnetic shear often remains even after a flare \citep[e.g.,][]{Wang1994}. Thus, the field opening of a CME may contribute to some increase in the open magnetic flux of the sun. Instead of searching for field opening events in all available CME images, we examine the monthly distribution of the CMEs  detected by {\it SOlar Heliospheric Observatory}'s (SOHO) Large Angle and Spectrometric COronagraph (LASCO) C3, the inner boundary of whose field of view is 3.7 solar radii, slightly above the usual source surface (1.6--3.25 solar radii).  

In this Letter, we study the dynamic behavior of the sun deviating from the sunspot cycle by investigating CHs and far-reaching CMEs, with a special attention paid to the declining phase of the cycle. For this purpose, we examine three indicators of coronal dynamic activities in solar cycle 23: (1) the areas of the polar and low latitude CHs, (2) the events of CH splitting and merging, and (3) CMEs observed by SOHO/LASCO C3. The sunspot number and the mean solar magnetic field will serve as reference ``magnetic'' activity levels to be compared with those ``dynamic'' activities.  In \S~\ref{sec:data}, the data we have used are described. In \S~\ref{sec:results}, we present the time characteristics of the solar coronal activity indicators throughout solar cycle 23 along with our interpretation of them. In \S~\ref{sec:discussion}, we provide a conclusion of our study.

\section{Data} \label{sec:data}

Our study deals with solar cycle 23, which started in 1996 May and ended in 2008 January. 
For the reference sunspot number to be compared with coronal activity indicators, we take the sunspot number data version 1.0 provided by the World Data Center-SILSO, Royal Observatory of Belgium, Brussels \citep{SIDC}. 
Since our study is concerned with the long-term variation of solar activity over the entire solar cycle 23, the 13 month smoothed monthly sunspot number (SSN) is adopted and presented. The maximum of the SSN in solar cycle 23 is located in 2000 April and its second peak in 2001 November. We will refer to the period between these two peaks as the sunspot maximum period of solar cycle 23, which may also be called the ``Gnevyshev gap in SSN.''  

For the mean solar magnetic field, we use the data generated and provided by the Wilcox Solar Observatory of the Stanford University \citep{Scherrer1977}.\footnote{\url{http://wso.stanford.edu/}} 
Since the mean solar magnetic field fluctuates a lot day by day, not only in its magnitude, but also in its sign, we here present the upper (positive) and lower (negative) envelopes of the data averaged over adjacent 27 days (Figure~1). 

As for the CH areas, we use the CH data of UCOHO (URSIgram code for CH information) produced by the Space Weather Prediction Center (SWPC) of the National Oceanic and Atmospheric Administration (NOAA). 
The UCOHO data are generated by examining solar images by human eye, and may be dependent on the subjectivity of the examiners. There is another set of CH area data, which is generated by the CH Automated Recognition and Monitoring (CHARM) algorithm \citep{Krista2009}. 
Although the CHARM data are free from human subjectivity, 
the areas of the polar CHs and the isolated low latitude CHs are not distinguished in them. Since the UCOHO data provide this distinction, we have used only the UCOHO data. 

To find the splitting and merging events of CHs, we have first examined the Solar Synoptic Maps, which are drawn by human hands and provided by 
NOAA/SWPC.\footnote{\url{http://www.swpc.noaa.gov/products/solar-synoptic-map}}
After we have identified probable candidates in the SWPC Solar Synoptic Maps, we have examined the {\it SOHO}  Extreme ultraviolet Imaging Telescope (EIT) 195~{\AA}  images and also checked the magnetic polarities in the {\it SOHO} Michelson Doppler Imager (MDI) magnetograms. We have selected only those events in which the continuous process of splitting or merging can be seen in consecutive days. 
Because the results can depend on human subjectivity, two of the authors (M.J. and S.H.) have independently performed the entire selection process and decided only on the events that both have identified without compromise. 

The data of CMEs detected by {\it SOHO}/LASCO C3 are retrieved from the online CME catalog.\footnote{\url{http://cdaw.gsfc.nasa.gov/CME_list/}} 
From 1996 May to 2007 December, 10,005 CMEs are detected by LASCO C3 while a total of 13,106 CMEs are detected by C2 or C3.

\begin{figure}[ht!]
\figurenum{1}
\plotone{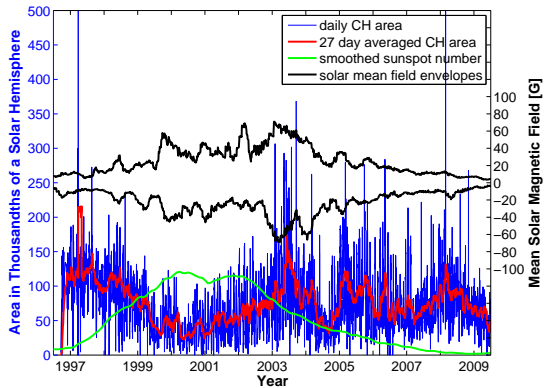}
\caption{Total CH area and the mean solar magnetic field along with SSN in solar cycle 23. The red line represents the 27 day averaged CH area in units of thousandths of a solar hemisphere area. The daily CH area is given in blue. The positive and negative envelopes of the mean solar magnetic field in Gauss averaged over 27 days are given in black. The SSN is presented in green.
\label{fig1}}
\end{figure}

\section{Results} \label{sec:results}

\subsection{Time Characteristics of Coronal Activity Indicators} \label{evol}

The daily CH area obtained from the UCOHO data and its 27 day averaged value in units of thousandths of a solar hemisphere area are plotted as a function of time in Figure~1 along with SSN and the envelopes of the 27 day averaged mean solar magnetic field. The positive maximum of the mean solar magnetic field envelope is located in 2002 November and its negative maximum is located in 2003 February, both at least a year after the second maximum of SSN. Although the maxima are located after the sunspot maximum period, the mean solar magnetic field is generally stronger in the sunspot maximum period than in the sunspot number declining phase. Therefore, the evolutionary trend of the mean solar magnetic field can be said to be more or less consistent with that of the sunspot number in spite of the delay of the former. 

The 27 day averaged CH area has the first maximum in 1997 March, five months after the sunspot minimum in 1996 October. Its second maximum is located in 2003 May, 18 months after the second maximum of SSN. If we consider the data before mid-2003, the CH area and the sunspot number are highly 
anti-correlated as argued in previous studies \citep{Bravo1994, Bravo1997, Dorotovic1996, Hesswebber2014}, which mostly considered polar CH areas. Now we have plotted the polar CH area and the isolated CH area separately in Figure~2. The polar CH area appears more or less anti-correlated with the sunspot number, whereas the isolated CH area is not so at all. It is to be noted that the polar CH area has a bumped distribution around the second maximum of the sunspot number for a duration of about 2.5 years. The distribution of the isolated CH area shows the maximum in 2004 January, 26 months after the second sunspot maximum, and the second maximum in 2005 March, and then declines almost following the sunspot number trend. 
\begin{figure}[ht!]
\figurenum{2}
\plotone{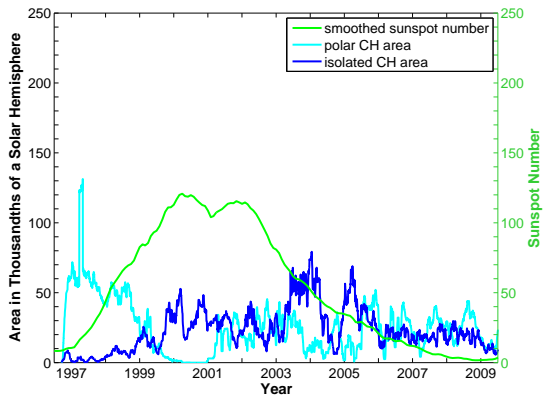}
\caption{Areas of the polar CHs and the isolated low latitude CHs along with SSN in solar cycle 23. The cyan line is for the polar CH area and the blue line for the isolated low latitude CH area. Both present 27 day averaged values in units of thousandths of a solar hemisphere area. The SSN is presented in green.
\label{fig2}}
\end{figure}

The numbers of splitting and merging events of CHs in each year of solar cycle 23 are given as a histogram in Figure~3. 
Although there are only 79 events in total, we clearly see that the events are dominantly populated in the declining phase of the sunspot cycle, and definitely avoid the sunspot maximum period. Especially in 2001, which is the local minimum between the two maxima in sunspot number, no splitting or merging events are found. 
A comparison of Figure~2 and Figure~3 suggests that the occurrence rate of CH splitting or merging in the declining phase more or less follows the variation of the isolated CH area, whereas the CH splitting or merging in the rising phase is seemingly related to the polar CHs. 
 Motivated by the proposition of \citet{Antiochos2007}, we have carefully looked for a CH totally nested in another CH, but have found none in solar cycle 23. 
\begin{figure}[ht!]
\figurenum{3}
\plotone{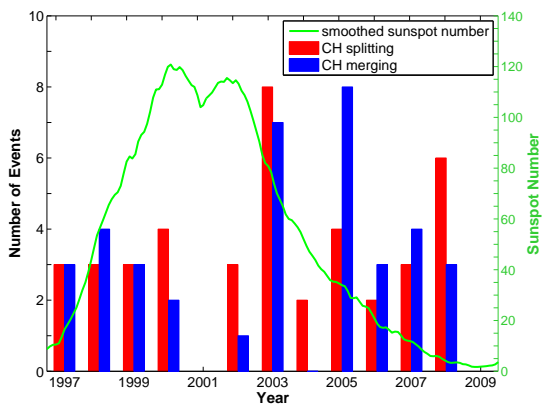}
\caption{Number of CH splitting and merging events in each year of solar cycle 23. The red histogram represents the annual number of CH splitting events, and the blue histogram represents that of CH merging events. The SSN is presented in green.
\label{fig3}}
\end{figure}

In Figure~4, we have plotted the monthly number of C3 CMEs from 1997 to 2009. 
Comparison with the sunspot number clearly shows that C3 CMEs are highly overpopulated in the declining phase of the solar cycle. The Gnevyshev gap in the CME distribution almost overlaps the Gnevyshev gap in sunspot number, but it is not so impressive as the peaks in 2005 and 2007, located much later than the second sunspot maximum in 2002. 
\begin{figure}[ht!]
\figurenum{4}
\plotone{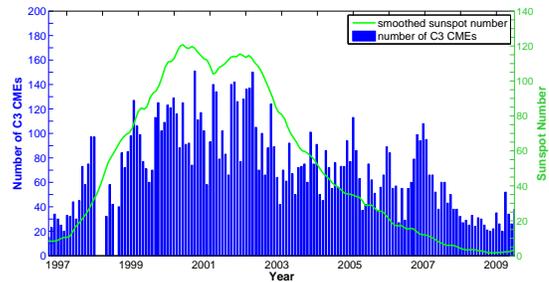}
\caption{Histogram of the LASCO C3 CMEs  in each month of solar cycle 23. The SSN is presented in green.
\label{fig4}}
\end{figure}

\subsection{Interpretation and Discussion} \label{interp}

To interpret the above results, we adopt a quasi-steady source surface model, which may not necessarily be a potential field. To better understand this model, we perform thought experiments as follows. 

In the PFSS model, an increase of the boundary normal field by a constant factor everywhere in the solar surface would not change the coronal field configuration, but the field strength only. The open flux through the source surface would be increased by the same factor, but the shape and area of the open field region would be conserved. Thus, the total CH area may be a rough proxy of the ratio of the open flux and the total flux (of either polarity) rather than a proxy of the open flux itself. 

Now we suppose that the total flux (of each polarity) through the solar surface be fixed while the normal field distribution there could change. In the PFSS model, a coronal magnetic field dominated by lower-order multipole moments (simpler photospheric field distributions) would have more open magnetic flux than one with higher-order multipole moments (more complex photospheric field distributions) because in the latter case, small-scale positive and negative polarity patches are closely located and they are connected by low-lying field lines below the source surface. 

From these two thought experiments, we infer that as the global solar magnetic field deviates more and more from a simple dipole configuration, the total CH area would be reduced in a PFSS model. However, this argument is not valid when electric currents exist in the corona. Although non-potential fields with a source surface boundary condition have been little studied \citep[e.g.,][]{Aly1993}, it is expected that we will have more open flux in the presence of stronger currents because a current-bearing magnetic field  generally tends to distend outward \citep{Mikic1994, Choe1996}. The more complex the field configuration is, the more coronal currents tend to develop.

The above argument can explain why the total CH area takes a maximum near the solar minimum (Figure 1), in which the dominance of the polar CHs (Figure 2) is indicative of a dipole-like global flux distribution. Near the solar maximum, the polar CH area becomes almost zero (Figure 2). In this stage, the polarity reversal takes place in the polar regions and the main CHs migrate to lower latitudes. In the early declining phase of the solar cycle (for about three years after the second sunspot maximum), the polar CHs revive, but the lower latitude CHs dominate over them in area (Figure 2). Thus, the surface flux distribution in this stage is far from a simple dipole. If the magnetic field were current-free (potential) in this stage, the open field area (total CH area) would be much smaller than near the solar minimum. In reality, however, the two area values are almost comparable (Figure 1). This implies that the magnetic fields in the declining phase bear much larger volume currents than in the rising phase and the solar maximum period. We speculate that these currents are generated by field-twisting/braiding motions, which are regarded as more horizontal than vertical because new flux emergence keeps decreasing in that stage.   

Here, we roughly differentiate the dynamic origins of solar activities into two: the flux emergence and the horizontal motion of magnetic fluxes, although one usually involves the other in reality. The flux emergence would be reflected above all in the sunspot number and also in the mean solar magnetic field in a more indirect way. The horizontal motions of magnetic fluxes would be reflected in splitting or merging of CHs as argued by \citet{Titov2011}. Our statistics suggest that horizontal motions of magnetic fluxes should be more activated in the declining phase of a solar cycle than in the sunspot maximum period when new flux emergence is at its maximum. In a crude picture, flux emergence would generate a thin current layer in the interface between new and extant fluxes and easily lead to flare-like reconnection events. A large-scale  horizontal motion in the declining phase would rather generate a large-scale volume current, which takes more  time until being released in a large-scale eruption \citep{Shibata2011}. It is thus understandable that  while the frequency of M- and X-class flare occurrences quite well follows the time evolution of SSN \citep{Hathaway2015}, C3 CMEs are highly overpopulated in the declining phase (Figure~4). 

To figure out how individual splitting or merging processes take place, we have  examined the $\rm H \alpha$ and white-light images of the sun provided by the Big Bear Solar Observatory\footnote{\url{http://www.bbso.njit.edu/Research/FDHA/}} and  compared them with the SWPC Solar Synoptic Maps. Out of the total 79 splitting or merging events, we have identified 11 events, which can be  associated with activities of nearby filaments or active regions, and listed them in Table~1. Most splitting events are related to the intrusion of a beta region, which contains sunspots of mixed polarities, as proposed by \citet{Titov2011}. Two splitting events are related to fading (S1) or disappearance (S4) of nearby filaments, and another (S2) to relocation (or new formation) of filaments between the split CHs. The three merging events are all related with disappearance of filaments between or near the merging CHs. One filament involved in event M2 is identified as having an eruption. 
It is plausible that a pseudo-streamer magnetic topology associated with a parasitic polarity between the merging CHs favored this eruption in a similar way as was modeled by \citet{Torok2011}. We were, however, not able to relate the activities of other filaments in the studied events with any observed eruption.

\begin{deluxetable}{cll}
\tabletypesize{\scriptsize}
\tablecaption{Splitting (S) or merging (M) events of CHs associated with 
activities of filaments or active regions nearby}
\tablewidth{8cm}
\tablehead{
\colhead{Event} & \colhead{Date} & \colhead{Activity} 
}
\startdata
S1 & 1998 Jun 6 &  Nearby filaments faded before the CH splitting. \\
S2 & 1998 Sep 19 &  Filaments located between the split CHs. \\
S3 & 2000 Mar 17 &  A beta region AR 8913 intruded. \\ 
S4 & 2000 Nov 9 &  A nearby filament  disappeared. \\
S5 & 2002 Feb 5 &  A beta region AR 9807 intruded. \\
S6 & 2003 Mar 12 &  A beta region AR 0311 intruded. \\
S7 & 2003 Jun 3 &  A beta region AR 0376 intruded. \\
S8 & 2003 Oct 17 &  A beta region AR 0481 intruded. \\
M1 & 1999 May 15  & A nearby filament disappeared. \\
M2 & 2003 May 8  & Two filaments between the CHs disappeared.  \\
M3 & 2005 Aug 30  & A filament between the CHs disappeared. \\
\enddata
\end{deluxetable}

\section{Conclusion} \label{sec:discussion}

In this Letter, we have investigated the solar dynamic activities related to CHs and far-reaching CMEs throughout solar cycle 23. While the total CH area is at maximum near the sunspot minimum, when polar CHs prevail, its second maximum lies 26 months after the second sunspot maximum, when the low latitude CHs dominate over polar CHs. Most events of CH splitting or merging are populated in the declining phase of the cycle and some in the rising phase, but very few between the two sunspot maxima. The splitting or merging events are attributed to the dynamic motions of magnetic fluxes in the solar surface \citep{Titov2011}. The LASCO C3 CMEs are highly overpopulated compared to sunspot numbers throughout the declining phase. Our results suggest that solar dynamic activities due to the horizontal motions of magnetic fluxes extend far in the declining phase of the sunspot cycle, even beyond a typical Gnevyshev gap. 

\acknowledgments

We thank the referee for constructive comments. 
We thank C.~C.~Balch and other forecasters at NOAA/SWPC for providing the UCOHO data for us to use.
This work was supported by Basic Science Research Program through the National Research Foundation of Korea (NRF) funded by the Ministry of Education (NRF-2013R1A1A2058937). 
The CME catalog is generated and maintained at the CDAW Data Center by NASA and the Catholic University of America in cooperation with the Naval Research Laboratory. 
SOHO is a project of international cooperation between ESA and NASA. 

\vspace{5mm}
\facilities{SOHO}


\listofchanges

\end{document}